# Reducing the Lift-Off Effect on Permeability Measurement for Magnetic Plates From Multifrequency Induction Data

Mingyang Lu, Wenqian Zhu, Liyuan Yin, Anthony J. Peyton, Wuliang Yin, *Senior Member, IEEE*, and Zhigang Qu

*Abstract*—Lift-off variation causes errors in eddy current measurement of nonmagnetic plates as well as magnetic plates. For nonmagnetic plates, previous work has been carried out to address the issue. In this paper, we follow a similar strategy, but try to reduce the lift-off effect on another index—zero-crossing frequency for magnetic plates. This modified index, termed as the compensated zero-crossing frequency, can be obtained from the measured multifrequency inductance spectral data using the algorithm we developed in this paper. Since the zero-crossing frequency can be compensated, the permeability of magnetic plates can finally be predicted by deriving the relation between the permeability and zero-crossing frequency from Dodd and Deeds method. We have derived the method through mathematical manipulation and verified it by both simulation and experimental data. The permeability error caused by liftoff can be reduced within 7.5%.

## I. INTRODUCTION

THE magnetic property (permeability) of metallic plates can be inferred by using both multifrequency and pulse eddy current testing methods. However, both methods suffer from errors caused by the so-called lift-off effect. To address this issue, a range of methods such as using different signal processing, feature extraction [1]–[5], sensor structure [6], [7], and detection principles [8]–[15] have been investigated by researchers. Multifrequency eddy current sensing in the context of nondestructive testing applications has been the focus of the authors' research in recent years. Conductivity and permeability depth profiling [16], [17], and noncontact microstructure monitoring [18]–[21] have been

Manuscript received February 6, 2017; revised April 11, 2017; accepted June 1, 2017. Date of publication October 24, 2017; date of current version December 7, 2017. This work was supported by the U.K. Engineering and Physical Sciences Research Council. The Associate Editor coordinating the review process was Dr. Sergey Kharkovsky. *(Corresponding author: Zhigang Qu.)*
M. Lu, W. Zhu, and A. J. Peyton are with the School of Electrical and Electronic Engineering, University of Manchester, Manchester M13 9PL, U.K.
L. Yin is with the School of Information Engineering and Automation, Kunming University of Science and Technology, Kunming 650093, China.
W. Yin is with the School of Electrical and Electronic Engineering, University of Manchester, Manchester M13 9PL, U.K., and also with the College of Electronic Information and Automation, Tianjin University of Science and Technology, Tianjin 300222, China (e-mail: wuliang.yin@manchester.ac.uk).
Z. Qu is with the College of Electronic Information and Automation, Tianjin University of Science and Technology, Tianjin 300222, China (e-mail: zhigangqu@tust.edu.cn).
Color versions of one or more of the figures in this paper are available online at http://ieeexplore.ieee.org.
Digital Object Identifier 10.1109/TIM.2017.2728338

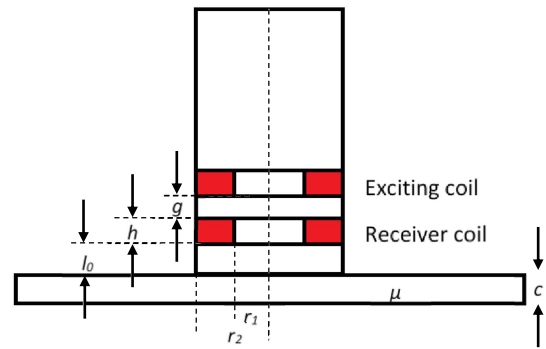

Fig. 1. Sensor configuration.

explored. Other colleagues in the field have also explored multifrequency non-destructive testing (NDT) for crack and defect detection [22]–[25]. In a recent paper [26], a new algorithm was proposed to compensate the lift-off effect for the nonmagnetic plate. Most of the aforementioned works are for nonmagnetic materials or use relatively simple phase features for magnetic plates. A feature called zero-crossing frequency, which is relatively robust to lift-off variation, was proposed by Zhu *et. al.* [27], [28], [29] and Peyton *et. al.* [30] for testing magnetic plates with applications in steel rolling production and rail inspection.

In this paper, we further extend the findings in [26]–[30] and consider a simple coil configuration (one transmitter and one co-axial receiver), but compensate the change of the zero-crossing frequency due to liftoff by exploiting a sophisticated algorithm which uses two fundamental facts. First, the zero-crossing frequency of the inductance spectral signal decreases with increased liftoff, and second, the overall magnitude of the signal decreases with increased liftoff. Since the relation between zero-crossing frequency and permeability was established previously as shown in [20], the compensated permeability is also independent of liftoff. Theoretical derivation, numerical simulation, and experiments show that both the compensated zero-crossing frequency and predicted permeability are nearly liftoff independent.

## II. SENSOR DESCRIPTION

The sensor is composed of two coaxially arranged coils, one as the transmitter and the other as the receiver, both of which have the same dimensions. A schematic plot of the sensor is shown in Fig. 1, with its dimensions in Table I. The design of this sensor is such that both the measurements and the analytical solution of Dodd and Deeds are accessible.

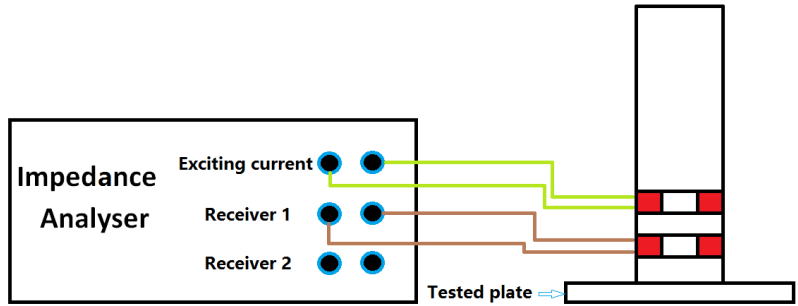

Fig. 2. Experimental wiring schematic. Illustrate how the sensor and instrument (Impedance Analyser SL 1260) were connected during the experiment process [26].

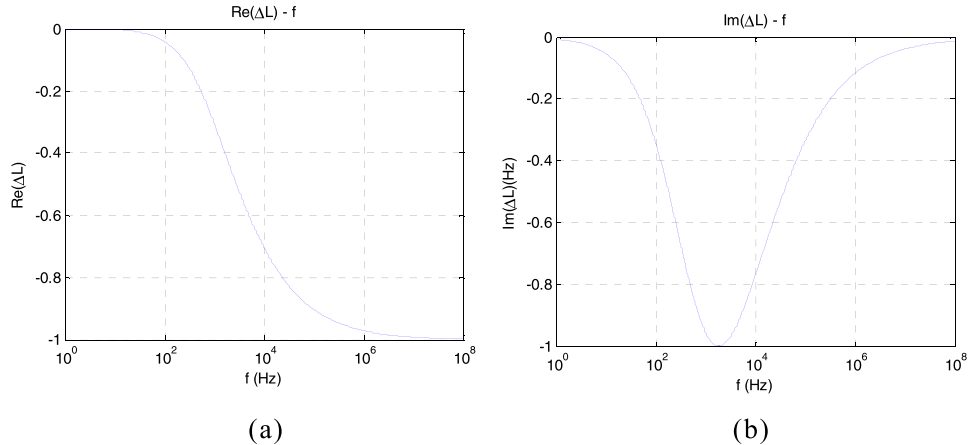

Fig. 3. Normalized diagram of inductance changes with frequency due to a nonmagnetic plate. (a) Real part. (b) Imaginary part.

TABLE I
COIL PARAMETERS

| | |
|---|---|
| $r_1$ | 11.4mm |
| $r_2$ | 12mm |
| $lo$ (lift-off) | 0.8mm |
| $h$ (height) | 1.5mm |
| $g$ (gap) | 1mm |
| Number of turns  N1 = N2 | 20 |

## III. ZERO-CROSSING FREQUENCY FEATURE

The behavior of the sensor system depends on the material's properties such as permeability under the inspection. In nonmagnetic and conductive materials, eddy currents are the main effect. However, the magnetic field produced by a multifrequency sensor acts on a magnetic plate in two ways [35]. First, it tends to magnetize the magnetic plate, which increases the coil system inductance. Second, the alternating current magnetic field also induces eddy currents in the magnetic plate, which tend to oppose the background or original driving current and reduce the coil system inductance. At lower frequencies, magnetization dominates and the $\Delta L$ as a result of the magnetic plate is positive. As the frequency is increased, the effects of eddy currents become more dominant and the $\Delta L$ decreases, at some point becoming negative, and eventually approaching a constant value at high frequencies. When the two effects are in balance there is a zero-crossing point frequency (i.e., the frequency at which the Re($\Delta L$) is zero). And normalized diagrams of $\Delta L - f$ for the sensor above nonmagnetic and magnetic materials were explored previously in [19] and are in Figs. 3 and 4. Here, $\Delta L$ denotes the mutual inductance changes between transmitter and receiver for different frequencies.

The physical meaning of zero-crossing frequency can also be interpreted in term of power transfer.

As can be seen from Fig. 4, the real part of the inductance shifts up as the relative permeability increases from 1 (nonmagnetic material). The relation between the impedance and inductance of the sensor is $Z = R + jwL$ (Here, the dc (direct current) resistance of coils is $R$ which is a constant value and is not considered for alternating current (ac) calculations) [7], [16], [20]. In addition, only the change in impedance is considered in this paper, which is not affected by $R$. Therefore, the real part of the inductance is in direct proportion to the imaginary part of the impedance, which is associated with system inductive energy storage. And the imaginary part of the inductance is proportional to the real part of the impedance, which can be physically explained as system resistance loss (or dissipative part). Therefore, the capability of the system storing inductive energy increases as the relative permeability. Due to the fact that eddy currents reduce inductive energy storage in the sensor, therefore there

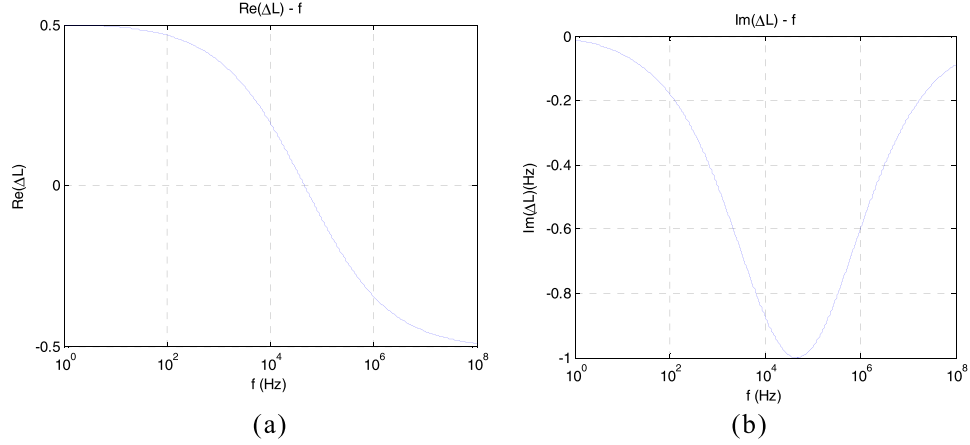

Fig. 4. Normalized diagram of inductance changes with frequency due to a magnetic plate. (a) Real part. (b) Imaginary part.

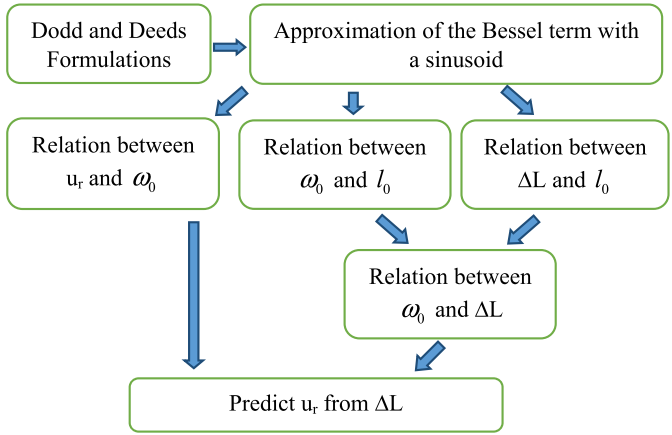

Fig. 5. Block diagram of the permeability prediction procedure.

exists a "zero" point where the system stores the same energy as it is in air.

## IV. THEORETICAL DERIVATION OF THE PREDICTED PERMEABILITY

Previously, we have observed that the zero-crossing frequency decreases with increased liftoff [7], [26]. It is also common knowledge that the signal amplitude also decreases with increased liftoff. Therefore, we hypothesize that an algorithm can be developed to compensate the variation in the zero-crossing frequency with the signal amplitude. In the following, we will derive such an algorithm. Some of the derivations are common to the nonmagnetic plate case [26], but for completeness, derivation process is provided in the appendix and main steps are summarized in Fig. 5. Where $\omega_0$ is the zero-crossing frequency, $l_0$ is the liftoff of the sensor, $\Delta L$ is the measured inductance changes, and $u_r$ is the material permeability.

Through mathematic manipulation, the compensated zero-crossing frequency can be obtained

$$\omega_0 = \frac{\mu_r \alpha_0^2}{\mu_0 \sigma} = \frac{\pi^2 \omega_1}{\left(\pi^2 + 4 \ln \frac{\Delta L_0}{\Delta L_m}\right)} \quad (1)$$

where $\Delta L_m$ denotes the magnitude of the inductance change with start point frequency (any frequency within the range of 1–10 Hz as shown in Fig. 6) or high frequency (1 MHz in Fig. 6 when the real part of inductance change is almost stable with frequency) under the lowest liftoff. (Here, the lowest liftoff is 0.8 mm in this paper.) While $\Delta L_0$ denotes magnitude of the inductance change under the current unknown liftoff, $\omega_1$ denotes the measured zero-crossing frequency.

Since the relation between permeability and zero-crossing frequency is introduced in the appendix, the permeability can be predicted in the following equation:

$$\mu_r = \frac{\mu_0 \sigma \omega_0}{\alpha_0^2} = \frac{\mu_0 \sigma \pi^2 \omega_1}{\alpha_0^2 \left(\pi^2 + 4 \ln \frac{\Delta L_0}{\Delta L_m}\right)}. \quad (2)$$

It can be seen in (6) that through a compensation scheme and using the knowledge of the permeability and the amplitude at a certain liftoff, the original permeability (permeability prior to introducing the liftoff $l_0$) can be recovered.

## V. SIMULATIONS AND EXPERIMENTS

Experiments and simulations were carried out to verify the performance of the compensation algorithm; the predicted permeability at different liftoffs was compared. Here, the real part of the inductance is defined from the mutual impedance of the transmitter and the receiver coils

$$\text{Im}(\Delta L) = \text{Im}\left(\frac{Z(f) - Z_{\text{air}}(f)}{j 2\pi f}\right)$$
$$= \text{Re}\left(\frac{-(Z(f) - Z_{\text{air}}(f))}{2\pi f}\right) \quad (3)$$
$$\text{Re}(\Delta L) = \text{Re}\left(\frac{Z(f) - Z_{\text{air}}(f)}{j 2\pi f}\right)$$
$$= \text{Im}\left(\frac{-(Z(f) - Z_{\text{air}}(f))}{2\pi f}\right) \quad (4)$$

where $Z(f)$ denotes the impedance of the coil with the presence of a metallic plate while $Z_{\text{air}}(f)$ is that of the coil in air.

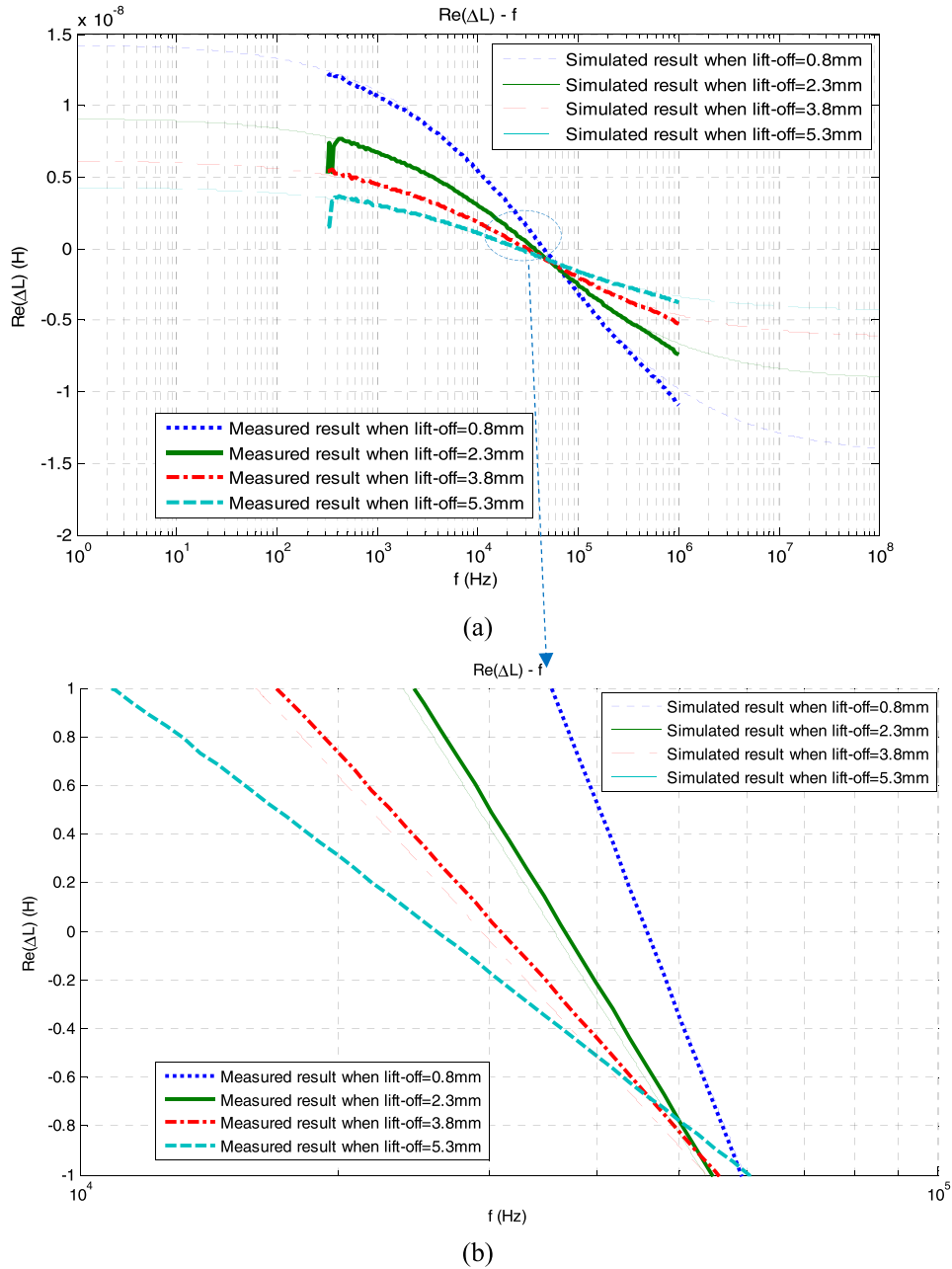

Fig. 6. (a) Simulated and measured real parts of $\Delta L$ for a plate with $u_r = 125.2$ at a range of liftoffs. (b) zoomed-in zero-crossing frequency part with a linear scale.

### A. Simulations

The sensor configuration used in simulations based on the Dodd and Deeds method is shown in Fig. 1. The simulated target is ferrous plate with a relative permeability of 125.2 and a conductivity of 6.624 MS/m under varying liftoffs of 0.8, 2.3, 2.8, 3.3, 3.8, 4.3, 4.8, and 5.3 mm. The simulations based on the Dodd and Deeds methods were realized by a custom solver developed using MATLAB. The solver can be used to calculate the Dodd and Deeds solution [(1)–(10) in the appendix] and the compensated zero-crossing frequency using (20)–(22) in the appendix. The solver can take a range of different parameters such as frequency, sample conductivity, permeability, and liftoff. In addition, the solver has been converted and packaged to an executable program.

### B. Experimental Setup

For the experimental setup, the metal plate used is composed of mixed ferrite and austenite.

The permeability of the plate ($u_r = 125.2$) was obtained by fitting the inductance–frequency curve measured by a commercial impedance analyzer and a well characterized cylindrical sensor to that simulated using the Dodd and Deeds method.

During the fitting, we did not find significant effect due to frequency dependent and complex-valued permeability. Of course, if materials do show frequency dependent and complexed value, then we need to take this into account, for example as in paper < Frequency-dependence of relative permeability in steel > by Bowler [34].

TABLE II
RELATIVE PERMEABILITY MEASUREMENTS FOR DIFFERENT LIFT OFFS

| Lift-off (mm) | Actual relative permeability | Relative permeability without compensation | Relative permeability calculated from Compensated zero-crossing frequency | Relative error for non-compensated permeability | Relative error for compensated permeability |
|---|---|---|---|---|---|
| 0.8 | 125.2 | 120.179393 | 120.179393 | 4.01% | 4.01% |
| 2.3 | 125.2 | 98.6480926 | 119.101695 | 21.21% | 4.87% |
| 2.8 | 125.2 | 89.1353189 | 118.210084 | 28.81% | 5.58% |
| 3.3 | 125.2 | 82.2564103 | 117.333333 | 34.30% | 6.28% |
| 3.8 | 125.2 | 76.104979 | 115.942384 | 39.21% | 7.39% |
| 4.3 | 125.2 | 72.5428571 | 116.286711 | 42.06% | 7.12% |
| 4.8 | 125.2 | 69.8668867 | 116.800415 | 44.20% | 6.71% |
| 5.3 | 125.2 | 68.6025918 | 117.259734 | 45.21% | 6.34% |

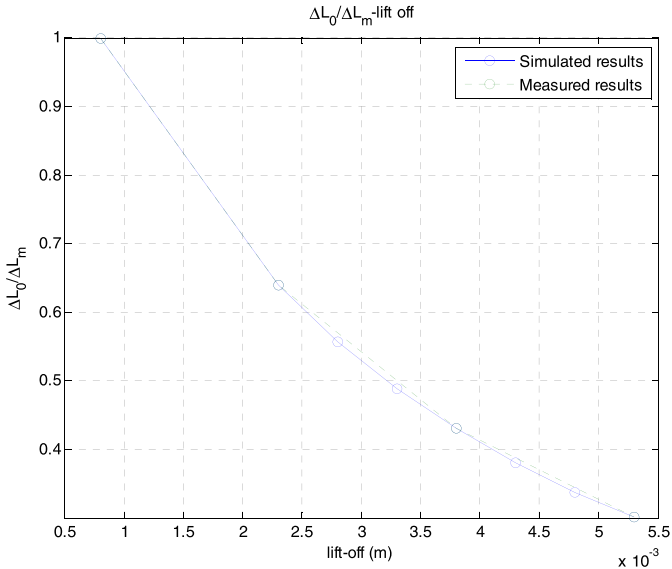

Fig. 7. Simulation and experimental results. The changes ratio of $\Delta L_0/\Delta L_m$ for a magnetic plate ($u_r = 125.2$) at a range of liftoffs.

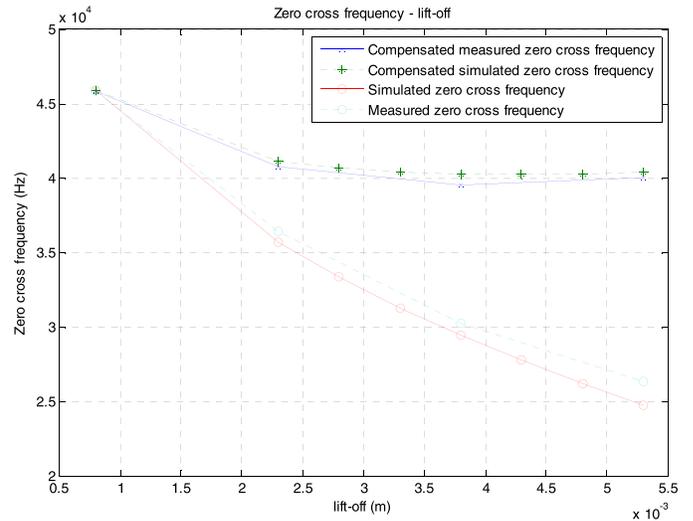

Fig. 8. Comparisons of as-measured (uncompensated) and compensated zero-crossing frequencies for a magnetic plate ($u_r = 125.2$) at a range of liftoffs.

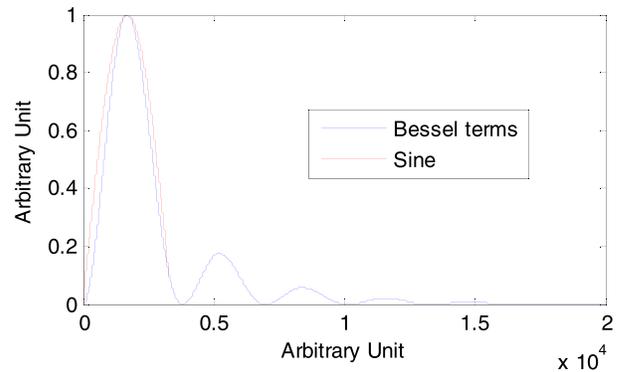

Fig. 9. Approximation of the Bessel term with a sinusoid.

The sensor configuration is the same as that used in simulations. And the multifrequency response of the sensor was obtained by a SL 1260 impedance analyzer with frequency sweeping mode. The frequency range of measurements is from 210 Hz to 1 MHz with 40 equally distributed samples in each decade step. SL 1260 impedance analyzer is currently the most powerful, accurate, and flexible frequency response analyser produced by SOLARTRON.

*C. Results*

It can be seen from Fig. 6 that the zero-crossing frequency decreases as liftoff increases. And at the same time, the magnitude of the signal decreases with increased liftoffs. The singularity points measured under low frequency are caused by the noise of SL 1260 impedance analyzer. In addition, the measured results deviate away from the simulated ones for frequencies higher than 500 kHz, which is due to the resonant phenomenon of the coil. (Its resonance frequency is about 2 MHz.)

As can be seen from Fig. 7, the changes ratio of $\Delta L_0/\Delta L_m$ decreases as liftoff increases, which can be used to compensate the drop in zero-crossing frequency with rising liftoffs. Where $\Delta L_m$ denotes the magnitude of the inductance change with start point frequency (any frequency within the range of 1–10 Hz as shown in Fig. 6) or high frequency. (1 MHz in Fig. 6 when the real part of inductance change is almost stable with frequency.) under the lowest liftoff (Here, the lowest liftoff is 0.8 mm in this paper.)

As can be seen from Fig. 8, the compensated zero-crossing frequency decreases slightly with initially increasing liftoff but remains almost constant for larger liftoffs, i.e., compensated zero-crossing frequency is virtually immune to lift-off variations for larger liftoffs. Both the compensated zero-crossing frequency and the zero-crossing frequency as measured for larger liftoffs are still lower than that for zero liftoff. Since the relative permeability is estimated according to (23) in the appendix, i.e., proportional to the zero-crossing frequency, the estimated relative permeability is smaller than the actual relative permeability. As can be seen from Table II, the estimated relative permeability calculated from the compensated zero-crossing frequency is much closer to the actual permeability than that without compensation.

## VI. CONCLUSION

This paper has proposed a compensation scheme for reducing the errors due to liftoff in estimating both the permeability and zero-crossing frequency from multifrequency eddy current measurements for magnetic plates. Previously, the zero-crossing frequency feature has been proven useful in predicting permeability and inferring steel microstructures. Based on the observation that the zero-crossing frequency decreases as liftoff increases and that the signal amplitude decreases with the increase of liftoff, an algorithm has been developed, which can compensate this variation in zero-crossing frequency and produce an index that is linked to relative permeability that can also virtually be independent of liftoff. Both the simulation and experimental results verified this. This is an important feature as lift-off variation is unavoidable in many practical applications.

A SL1260 impedance analyzer working in a swept frequency mode was used to acquire the multifrequency data in this paper. However, multifrequency impedances can also be abstracted simultaneously using composite multisine waveform excitation as in [23] or [34], which may improve the acquisition speed and calculation burden.

## APPENDIX

The Dodd and Deeds analytical solution describe the inductance change of an air-core coil caused by a layer of the metallic plate for both nonmagnetic and magnetic cases [21]. Another similar formula exists [22]. The difference in the complex inductance is $\Delta L(\omega) = L(\omega) - L_A(\omega)$, where the coil inductance above a plate is $L(\omega)$, and $L_A(\omega)$ is the inductance in free space.

The formulas of Dodd and Deeds are

$$\Delta L(\omega) = K \int_0^\infty \frac{P^2(\alpha)}{\alpha^6} A(\alpha)\phi(\alpha) d\alpha \tag{5}$$

where

$$\alpha_1 = \sqrt{\alpha^2 + j\omega\sigma\mu_r\mu_0} \tag{6}$$

$$\phi(\alpha) = \frac{(\mu_r\alpha - \alpha_1)}{(\mu_r\alpha + \alpha_1)} = \frac{\mu_r\alpha - \sqrt{\alpha^2 + j\omega\sigma\mu_r\mu_0}}{\mu_r\alpha + \sqrt{\alpha^2 + j\omega\sigma\mu_r\mu_0}}$$

$$= \frac{1 - \sqrt{1/\mu_r^2 + j\omega\sigma\mu_0/\mu_r\alpha^2}}{1 + \sqrt{1/\mu_r^2 + j\omega\sigma\mu_0/\mu_r\alpha^2}} \tag{7}$$

$$K = \frac{\pi\mu_0 N^2}{h^2(r_1 - r_2)^2} \tag{8}$$

$$P(\alpha) = \int_{\alpha r_1}^{\alpha r_2} x J_1(x) dx \tag{9}$$

$$A(\alpha) = e^{-\alpha(2l_0+h+g)}(e^{-2\alpha h} + 1) \tag{10}$$

where $\mu_0$ denotes the permeability of free space. $\mu_r$ denotes the relative permeability of plate. $N$ denotes the number of turns in the coil; $r_1$ and $r_2$ denote the inner and outer radii of the coil; while $l_0$ and $h$ denote the liftoff and the height of the coil, $g$ denote the gap between the exciting coil and receiver coil.

Equations (1)–(10) can be approximated based on the fact that $\phi(\alpha)$ varies slowly with $\alpha$ compared to the rest of the integrand, which reaches its maximum at a characteristic spatial frequency $\alpha_0$. The approximation is to evaluate $\phi(\alpha)$ at $\alpha_0$ and take it outside of the integral

$$\Delta L(\omega) = \phi(\alpha_0)\Delta L_0 \tag{11}$$

where

$$\phi(\alpha_0) = \frac{1 - \sqrt{1/\mu_r^2 + j\omega\sigma\mu_0/\mu_r\alpha_0^2}}{1 + \sqrt{1/\mu_r^2 + j\omega\sigma\mu_0/\mu_r\alpha_0^2}}. \tag{12}$$

Neglect $1/\mu_r^2$ in the following equation:

$$\phi(\alpha_0) = \frac{1 - \sqrt{j\omega\sigma\mu_0/\mu_r\alpha_0^2}}{1 + \sqrt{j\omega\sigma\mu_0/\mu_r\alpha_0^2}}$$

$$\Delta L_0 = K \int \frac{P^2(\alpha)}{\alpha^6} A(\alpha) d\alpha. \tag{13}$$

Note that in (11), the sensor phase signature is solely determined by $\phi(\alpha_0)$, which includes conductivity, permeability and $\alpha_0$. $\Delta L_0$ is the overall magnitude of the signal, which is strongly dependent on the coil geometrical parameters but independent of electromagnetic properties of the plate

$$\text{Assigning } \omega_1 = \frac{\mu_r\alpha_0^2}{\mu_0\sigma}. \tag{14}$$

Equation (13) can be expressed as

$$\phi(\alpha_0) = \frac{1 - \sqrt{j\omega/\omega_1}}{1 + \sqrt{j\omega/\omega_1}}. \tag{15}$$

In (11), as the real part of $\phi(\alpha_0)$ equals zero when $\omega = \omega_1$ and $\phi(\alpha_0)$ determines the phase of $\Delta L_0$, it can be seen that the zero-crossing frequency for the first order system is approximately $\omega_1$, and from (10) it is concluded that the zero-crossing frequency increases with $\alpha_0$.

Suppose a lift-off variation of $l_0$ is introduced, from (10), we can see that an increase of $l_0$ in liftoff is equivalent to multiplying a factor $e^{-2\alpha l_0}$

$$A(\alpha) = e^{-2\alpha l_0} e^{-\alpha(h+g)}(e^{-2\alpha h} + 1). \tag{16}$$

Due to the fact that $\Delta L_0 = K \int (P^2(\alpha)/\alpha^6) A(\alpha) d\alpha$ reaches its maximum at $\alpha_0$ and that the squared Bessel term $P^2(\alpha)$ is the main contributor, a simple function $\sin^2((\alpha\pi/2\alpha_0))$ with its maximum at $\alpha_0$ is used to approximate $\Delta L_0$

$$\Delta L_0 \approx \Delta L_m e^{-2\alpha l_0} \sin^2\left(\frac{\alpha\pi}{2\alpha_0}\right) \tag{17}$$

where $\Delta L_m$ denotes the magnitude of the inductance change with start point frequency (any frequency within the range of 1–10 Hz as shown in Fig. 6) or high frequency (1 MHz in Fig. 6 when the real part of inductance change is almost stable with frequency) when the lift-off is zero. While $\Delta L_0$ denotes magnitude of the inductance change under the current unknown liftoff. The approximation is appropriate in this context as the area covered by the Bessel terms and the Sine term are within 5% and it made the derivation of an analytical algorithm possible, as shown in Fig. 9 [26].

Equation (17) is then applied to obtain an analytical solution for $\alpha_0$.

The shift in $\alpha_0$ due to the effect of liftoff can be predicted as follows.

The new $\alpha$ should maximize $e^{-2\alpha l_0} \sin^2(\alpha\pi/2\alpha_0)$ and therefore $e^{-\alpha l_0} \sin(\alpha\pi/2\alpha_0)$.

The maximum can be obtained by finding the stationary point for $e^{-\alpha l_0} \sin(\alpha\pi/2\alpha_0)$.

Let $(e^{-\alpha l_0} \sin(\alpha\pi/2\alpha_0))' = -l_0 \cdot e^{-\alpha l_0} \sin(\alpha\pi/2\alpha_0) + \frac{\pi}{2\alpha_0} e^{-\alpha l_0} \cos(\alpha\pi/2\alpha_0) = 0$.

And through some mathematical manipulations, a new equation can be obtained

$$\frac{\alpha\pi}{2\alpha_0} = \tan^{-1}\left(\frac{\pi}{2\alpha_0 l_0}\right).$$

With small lift-off variation, $\alpha_0 l_0 \ll 1$ holds, the right side can be approximated as $(\pi/2) - (2\alpha_0 l_0/\pi)$.

Therefore, the revised $\alpha_{0r}$ is

$$\alpha_{0r} = \alpha_0 - \frac{4\alpha_0^2 l_0}{\pi^2}. \tag{18}$$

Combining (10) with (14), $\omega_1$ becomes

$$\omega_1 = \frac{(\alpha_0^2 \pi^4 - 8\pi^2 \alpha_0^3 l_0 + 16\alpha_0^4 l_0^2)\mu_r}{\pi^4 \sigma \mu_0}. \tag{19}$$

Combining (17) with (14), $\Delta L_0$ becomes

$$\Delta L_0 = \Delta L_m e^{-2\left(\alpha_0 - \frac{4\alpha_0^2 l_0}{\pi^2}\right)l_0} \cos^2\left(\frac{2\alpha_0 l_0}{\pi}\right)$$
$$= \Delta L_m e^{-2\left(\alpha_0 - \frac{4\alpha_0^2 l_0}{\pi^2}\right)l_0} \left(\frac{\cos\left(\frac{4\alpha_0 l_0}{\pi}\right) + 1}{2}\right).$$

Considering $\alpha_0 l_0 \ll 1$ and based on small-angle approximation $\cos(\theta) \approx 1 - (\theta^2/2)$, $\cos(4\alpha_0 l_0/\pi)$ is substituted with $1 - ((4\alpha_0 l_0/\pi)^2/2)$.

$\Delta L_0$ becomes

$$\Delta L_0 = \Delta L_m e^{-2\left(\alpha_0 - \frac{4\alpha_0^2 l_0}{\pi^2}\right)l_0} \left(1 - \frac{4\alpha_0^2 l_0^2}{\pi^2}\right).$$

Substituting $(1 - (4\alpha_0^2 l_0^2/\pi^2))$ with $e^{-(4\alpha_0^2 l_0^2/\pi^2)}$

$$\Delta L_0 = \Delta L_m e^{-2\left(\alpha_0 - \frac{4\alpha_0^2 l_0}{\pi^2}\right)l_0} e^{-\frac{4\alpha_0^2 l_0^2}{\pi^2}}$$
$$= \Delta L_m e^{-2\left(\alpha_0 - \frac{2\alpha_0^2 l_0}{\pi^2}\right)l_0}. \tag{20}$$

Taking natural logarithmic operation of both sides, we arrive at

$$\ln\frac{\Delta L_0}{\Delta L_m} = -2\left(\alpha_0 - \frac{2\alpha_0^2 l_0}{\pi^2}\right)l_0. \tag{21}$$

And further

$$4\alpha_0^2 l_0^2 - 2\pi^2 \alpha_0 l_0 - \pi^2 \ln\frac{\Delta L_0}{\Delta L_m} = 0.$$

This is now a quadratic equation with $\alpha_0 l_0$ as its variable. Therefore, the solution for $\alpha_0 l_0$ is

$$\alpha_0 l_0 = \frac{\pi^2 - \sqrt{\pi^4 + 4\pi^2 \ln\frac{\Delta L_0}{\Delta L_m}}}{4}. \tag{22}$$

The other solution $\alpha_0 l_0 = \pi^2 + (\pi^4 + 4\pi^2 \ln(\Delta L_0/\Delta L_m))^{1/2}/4$ does not satisfy the small lift-off condition $\alpha_0 l_0 \ll 1$ and therefore are discarded.

From (18), liftoff can be estimated as

$$l_0 = \frac{\pi^2 - \sqrt{\pi^4 + 4\pi^2 \ln\frac{\Delta L_0}{\Delta L_m}}}{4\alpha_0}. \tag{23}$$

Combining (15) with (19), the zero-crossing frequency with a liftoff of $l_0$ becomes

$$\omega_1 = \frac{\alpha_0^2 \left(\pi^2 + 4\ln\frac{\Delta L_0}{\Delta L_m}\right)\mu_r}{\pi^2 \sigma \mu_0}. \tag{24}$$

Equation (20) becomes a quadratic equation with an unknown $\alpha_0$

$$\alpha_0^2 \left(\pi^2 + 4\ln\frac{\Delta L_0}{\Delta L_m}\right)\mu_r - \pi^2 \sigma \mu_0 \omega_1 = 0.$$

And the solution is

$$\alpha_0 = \sqrt{\frac{\pi^2 \sigma \mu_0 \omega_1}{\left(\pi^2 + 4\ln\frac{\Delta L_0}{\Delta L_m}\right)\mu_r}}. \tag{25}$$

Therefore, the original zero-crossing frequency (zero-crossing frequency prior to introducing the liftoff $l_0$) can be obtained by combining (10) with (21)

$$\omega_0 = \frac{\mu_r \alpha_0^2}{\mu_0 \sigma} = \frac{\pi^2 \omega_1}{\left(\pi^2 + 4\ln\frac{\Delta L_0}{\Delta L_m}\right)}. \tag{26}$$

So the relative permeability reduces to

$$\mu_r = \frac{\mu_0 \sigma \omega_0}{\alpha_0^2} = \frac{\mu_0 \sigma \pi^2 \omega_1}{\alpha_0^2 \left(\pi^2 + 4\ln\frac{\Delta L_0}{\Delta L_m}\right)}. \tag{27}$$

It can be seen in (22) that through a compensation scheme and using the knowledge of the zero-crossing frequency and the amplitude at a certain liftoff, the original zero-crossing frequency (zero-crossing frequency prior to introducing the liftoff $l_0$) can be recovered.

ACKNOWLEDGMENTS

The authors would like to thank the UK Engineering and Physical Sciences Research Council (EPSRC) for their financial support of this research.

**Mingyang Lu** is currently pursuing the Ph.D. degree with the School of Electrical and Electronic Engineering, The University of Manchester, Manchester, U.K., focusing on developing a finite-element method model to solve electromagnetic simulation taking into account random geometry and material properties (including microstructure).

His current research interests include developing software to increase the efficiency of simulations to avoid remeshing, for example to consider a moving sensor as a moving effective field above a flaw (nondestructive testing application).

**Wenqian Zhu**, photograph and biography not available at the time of publication.

**Liyuan Yin**, photograph and biography not available at the time of publication.

**Anthony J. Peyton** received the B.Sc. degree in electrical engineering and electronics and the Ph.D. degree from the University of Manchester Institute of Science and Technology (UMIST), Manchester, U.K., in 1983 and 1986, respectively.

He was a Principal Engineer with Kratos Analytical Ltd., Manchester, from 1986 to 1989, where he was involved in developing precision electronic instrumentation systems for magnetic sector and quadrupole mass spectrometers. He joined the Process Tomography Group, UMIST, where he was a lecturer. In 1996, he was a Senior Lecturer with Lancaster University, Lancaster, U.K., where he was a reader in electronic instrumentation in 2001, and a Professor in 2004. Since 2004, he has been a Professor of electromagnetic tomography engineering with The University of Manchester, Manchester. His current research interests include instrumentation, applied sensor systems, and electromagnetics.

**Wuliang Yin** (M'05–SM'06) was a MT sponsored Senior Lecturer with the School of Electrical and Electronic Engineering, The University of Manchester, Manchester, U.K., in 2012, and became a Senior Lecturer in 2016. He has authored or co-authored of a book, more than 140 papers, and holds more than ten patents in the area of electromagnetic (EM) sensing and imaging. He currently leads several major grants from U.K. government bodies including TSB, EPSRC and is involved in several EU projects.

Dr. Yin was a recipient of the 2014 and 2015 Williams Award from the Institute of Materials, Minerals and Mining for his contribution in applying EM imaging in the steel industry and the Science and Technology Award from the Chinese Ministry of Education in 2000.

**Zhigang Qu** is currently with the College of Electronic Information and Automation, Tianjin University of Science and Technology, Tianjin, China.